\documentclass[conference, a4paper]{IEEEtran}
\IEEEoverridecommandlockouts
\usepackage{caption}
\usepackage{subfig}
\usepackage{graphicx}
\usepackage{tabularx}
\usepackage{amsmath,amssymb}
\usepackage{cite}
\usepackage[utf8]{inputenc}
\usepackage{import}
\usepackage{pgf,tikz}
\usepackage{dblfloatfix}
\usepackage{balance}
	\usepackage{fancyhdr}
\usepackage{geometry}
\ifCLASSINFOpdf
\else
\fi
\author{\IEEEauthorblockN{Quentin Bodinier,
		Faouzi Bader, and
		Jacques Palicot} 
	\IEEEauthorblockA{SCEE/IETR - CentraleSupélec, Rennes, France, \\}
	\IEEEauthorblockA{Email : \{firstname.lastname\}@supelec.fr}\vspace{-10pt}}

\newcounter{tmpeqcnt}
\begin{document}
	\DeclareGraphicsExtensions{eps}
	\graphicspath{{fig/}}
	\bstctlcite{IEEEexample:BSTcontrol}
	%
	\title{Coexistence in 5G: Analysis of Cross-Interference between OFDM/OQAM and Legacy Users}

	\maketitle

	\begin{abstract}
		To optimize the use of the spectrum, it is expected that the next generation of wireless networks (5G) will enable coexistence of newly introduced services with legacy cellular networks. These new services, like Device-To-Device (D2D) communication, should require limited synchronization with the legacy cell to limit the amount of signaling overhead in the network. However, it is known that  Cyclic Prefix-Orthogonal Frequency Division Multiplexing (CP-OFDM) used in Long Term Evolution-Advanced (LTE-A) is not fit for asynchronous environments. This has motivated the search for a new waveform, able to enhance coexistence with CP-OFDM. Namely, it has been widely suggested that new devices could use OFDM/Offset-Quadrature Amplitude Modulation (OFDM/OQAM) to reduce the interference they inject to legacy cellular users. However, values of interference are usually measured at the input antenna of the receiver, based on the PSD of the interfering signal. We showed in previous works that this measurement is not representative of the actual interference that is seen after the demodulation operations. Building on this finding, we provide in this paper the first exact closed forms of cross-interference between OFDM/OQAM and CP-OFDM users. Our results prove that using OFDM/OQAM only marginally reduces interference to legacy users, in contradiction with many results in the literature.

		
	\end{abstract}
	

	%
	\IEEEpeerreviewmaketitle

	\section{Introduction}
	\label{sec:intro}
	The advent of the 5th Generation of wireless communication
systems (5G) is envisioned to bring flexibility to cellular networks.
New services as Device-To-Device (D2D) or Machine-
To-Machine (M2M) communications are expected to be massively
deployed in the near future. Such new communication
devices have to coexist with incumbent legacy systems in
the cell, i.e. Long-Term-Evolution Advanced (LTE-A) users.
In such heterogeneous environments, perfect synchronization
between the different types of systems is not feasible. This loss
of synchronization will cause harmful interference between
active users, which will in turn degrade the overall system
performance.
This hurdle can be overcome through the design of new
waveforms that are robust against asynchronism, and well
localized in frequency. As a matter of fact, as soon as
the orthogonality between CP-OFDM users is destroyed, for
example because of the coexistence of unsynchronized
incumbent and secondary systems, their performance shrinks
dramatically \cite{Medjahdi2011}. This is mainly due to the fact that CP-OFDM
systems filter symbols with a time-rectangular window,
which causes poor frequency localization \cite{Farhang-Boroujeny2011, Baltar2007, Mahmoud2009} and high
asynchronism sensitivity in the multi-user context \cite{Mahmoud2009, Raghunath2009, Aminjavaheri2015}.

OFDM with Offset Quadrature Amplitude Modulation
(OFDM/OQAM) \cite{Farhang-Boroujeny2011,Bellanger2010}, is one of the main new waveform
schemes explored by the research community. Indeed, it overcomes
the cited CP-OFDM limitations and enables both higher
flexibility and reduction of interference leakage for multistandard
systems coexistence  \cite{Farhang-Boroujeny2011, Bellanger2010}. The main selling point
of OFDM/OQAM lies in its improved spectral containment
that is obtained through the filtering of each subcarrier with
a highly selective prototype filter. In this paper, we will study the case of OFDM/OQAM systems using the PHYDYAS
filter \cite{Bellanger2010}. Thanks to the enhanced spectral localization of the latter, OFDM/OQAM boasts
hardly measurable out-of-band (OOB) emissions, as its Power
Spectral Density (PSD) rapidly decreases below the ambient
noise level.

\begin{figure*}[!ht]
	\centering
	\begin{tikzpicture}[scale=0.9]
	\node (I) at (0,0) {};
	\node[draw,text width=2cm,align=center] (Q) at (3,0) {QAM Encoding};
	\node[draw,text width=2cm,align=center] (O) at (6.5,0) {OFDM Modulation};
	\node[draw,text width=2.1cm,align=center] (OD) at (14.5,0) {OFDM Demodulation};
	\draw[->,>=latex] (I) -- node[above]{$\ldots101\ldots$} (Q);
	\draw[->,>=latex] (Q) -- node[above]{$\mathbf{d}_\textbf{i}$} (O);
	\draw[->,>=latex] (10.25,0) -- node[above]{$y_\textbf{i}(t)$} (OD);
	\draw[gray,very thin, dotted](0, -1) -- (\linewidth,-1);
	\draw (10,0) circle (0.25); 
	\draw (10, 0.2) -- (10, -0.2);
	\draw (9.8, 0) -- (10.2, 0);
	\draw[->,>=latex] (O) -- node[above, pos = 0.8]{${s}_\textbf{i}(t)$} (9.75,0);
	\draw[->,>=latex, dashed] (O.east) -- node[right,pos=0.8]{$\ {s}_\textbf{i}(t)$} (10,-1.75);
	\draw[->,>=latex] (10,0.75) -- node[pos=0.2,right]{$w_\textbf{i}(t)$} (10,0.25);
	\draw[->,>=latex] (OD) -- node[above,pos = 0.3]{$\mathbf{\hat{d}}_\textbf{i}$} (\linewidth,0);
	\node (I2) at (0,-2) {};
	\node[draw,text width=2cm,align=center] (OQ) at (3,-2) {OQAM Encoding};
	\node[draw,text width=2.2cm,align=center] (F) at (6.5,-2) {OFDM/OQAM Modulation};
	\node[draw,text width=2.2cm,align=center] (FD) at (14.5,-2) {OFDM/OQAM Demodulation};
	\draw[->,>=latex] (I2) -- node[above]{$\ldots101\ldots$} (OQ)  ;
	\draw[->,>=latex] (OQ) -- node[above]{$\mathbf{d}_\textbf{s}$} (F);
	\draw[->,>=latex] (10.25,-2) -- node[below]{$y_\textbf{s}(t)$} (FD);
	\draw (10,-2) circle (0.25); 
	\draw (10, -2.2) -- (10, -1.8);
	\draw (9.8, -2) -- (10.2, -2);
	\draw[->,>=latex] (F) -- node[left, below, pos=0.8]{$s_\textbf{s}(t)$}  (9.75,-2);
	\draw[->,>=latex] (10,-2.75) -- node[pos=0.2,right]{$w_\textbf{s}(t)$}  (10,-2.25);
	\draw[->,>=latex, dashed] (F.east) -- node[right,pos=0.8]{$\ {s}_\textbf{s}(t)$} (10,-0.25);
	\draw[->,>=latex] (FD) -- node[above,pos = 0.3]{$\mathbf{\hat{d}}_\textbf{s}$} (\linewidth,-2);
	\node[above right,text width=1.5cm] (Inc) at (0, -1) {\textcolor{blue}{Incumbent}};
	\node[below right,text width=1.5cm] (Sec) at (0, -1) {\textcolor{orange}{Secondary}};
	\end{tikzpicture}
	\vspace{-5pt}
	\caption{Considered system model. Interference signals are marked with dashed arrows. 
	}
	\label{fig:scenario}
	\vspace{-20pt}
\end{figure*}
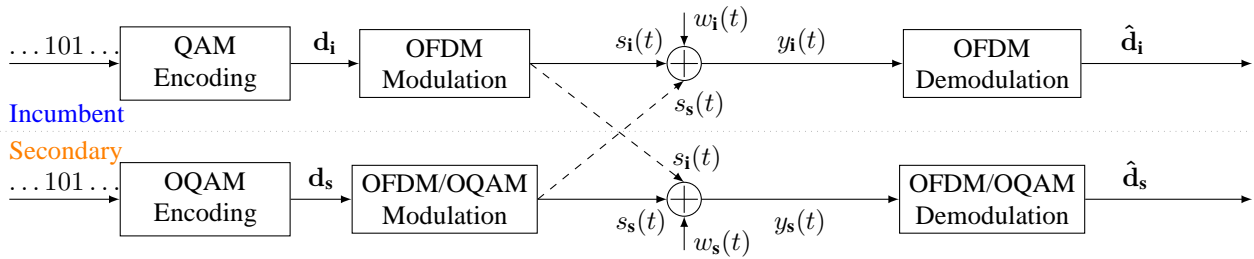

Building on this, a number of papers, for example
\cite{shaat2010computationally,skrzypczak2012ofdm, Ihalainen2008, Lin2015b}, have suggested to use OFDM/OQAM to coexist
efficiently with CP-OFDM based networks. All the studies on
that matter use the PSD-based model, originally proposed in
\cite{weiss04}, to rate the cross-interference between the OFDM/OQAM
and CP-OFDM systems. This model consists in integrating the
PSD of the interfering signal on each subcarrier of the user
that suffers from interference. Because of the advantageous
PSD properties of OFDM/OQAM, studies using the PSD-based
model predicted that using this waveform instead of
CP-OFDM for coexistence with LTE-A users would be highly
beneficial.

However, the expected gains were not confirmed
through simulations \cite{BodinierICC2016, Bodinier2016ICT}. 
Moreover, in \cite{Bodinier2016ICT}, we explained in a qualitative
manner why the PSD-based model was not fit to properly
estimate the cross-interference that is seen after the demodulation operations at both the receivers of OFDM/OQAM and
CP-OFDM systems. Furthermore, we showed through numerical simulations that the gains expected by
following the PSD-based model were highly overestimated.
However, a thorough mathematical analysis of the post-demodulation cross-interference
arising between coexisting OFDM/OQAM and
CP-OFDM systems is still lacking. This paper aims at resolving
this issue by providing mathematical closed forms of cross-interference
injected by OFDM/OQAM onto CP-OFDM and
vice-versa. The provided closed forms prove that OFDM/OQAM fails to protect incumbent legacy CP-OFDM users.

The remainder of this paper is organized as follows:
Section II presents the system model and a short overview
on CP-OFDM and OFDM/OQAM waveforms. In Section III,
the analysis of cross-interference is led. In Section IV, the
validity of the derived closed forms is asserted by comparison
with numerical simulations, and conclusions are given in Section V.

\textit{Notations:} scalars are noted $x$,vectors are bold-faced as $\mathbf{x}$, and ensembles are represented by a calligraphy letter $\mathcal{X}$. $n$ is the discrete symbol index, $m$ indexes subcarriers and $t$ is the continuous time. $\ast$ is the convolution operation, $x^*$ is the complex conjugate of $x$, $E_\alpha\{\}$ is the mathematical expectation with regards to the random variable $\alpha$ and $\mathcal{R}\{x\}$ is the real part of $x$.
	
	\section{Background and System Model}
	\label{sec:model}
	\begin{figure*}[!b]
		\vspace*{-15pt}
	\hrulefill
	\normalsize
	\setcounter{tmpeqcnt}{\value{equation}}
	\setcounter{equation}{17}
	\begin{equation}
	\mathbf{\eta}_{m_\textbf{s}\rightarrow{m_\textbf{i}}}[n_\textbf{i}] = \frac{1}{\sqrt{T}} \sum\limits_{n_\textbf{s}\in\mathbb{Z}}(-1)^{m_\textbf{s}n_\textbf{s}}\mathbf{d}_{m_\textbf{s}}[n_\textbf{s}]\mathbf{\theta}_{m_\textbf{s}}[n_\textbf{s}]\int\limits_{n_\textbf{i}(T+T_\text{CP})}^{n_\textbf{i}(T+T_\text{CP})+T}g\left(t-n_\textbf{s}\frac{T}{2}\right)e^{j2\pi (m_\textbf{s}
		-m_\textbf{i})\frac{t}{T}}dt, \forall n_\textbf{i} \in \mathbb{Z}. \label{eq:etastoidev}
	\end{equation}
	\setcounter{equation}{\value{tmpeqcnt}}
\end{figure*}

\subsection{Analyzed Coexistence Scenario}

\begin{figure}[width=0.9\linewidth]
	\centering
	\begin{tikzpicture}[scale=0.9]
	 \draw[->,>=latex] (0,0) -- node[at end, below left]{$\frac{f}{\Delta \text{F}}$} (\linewidth,0);
	 \draw[xstep=0.3,ystep=100,very thin] (0.3,0) grid (\linewidth*0.85,0.2);
	 \node[below] at  (\linewidth*0.85, 0) {$\frac{M}{2}-1$};
	 \node[below] at  (0.2, 0) {$-\frac{M}{2}$};
	 \draw[->,>=latex] (\linewidth*0.436, -0.3) -- node[at start,left]{$0$} (\linewidth*0.436, 1.5);
	 \draw[dashed,blue] (0.75, 0) -- (0.75, 1.5);
	 \draw[dashed,blue] (3.15, 0) -- (3.15, 1.5);
	 \draw[dashed,orange] (4.05, 0) -- (4.05, 1.5);
	 \draw[dashed,orange] (6.15, 0) -- (6.15, 1.5);
	 \node[blue] at (1.9,1.2) {$\mathcal{M}_\textbf{i}$}; 
	 \node[orange] at (5.1,1.2) {$\mathcal{M}_\textbf{s}$};
	 \draw[xstep=0.3, shift={(-0.15,0)},blue](0.9,0) grid (3.3,1);
	 \draw[xstep=0.3, shift={(-0.15,0)},orange](4.2,0) grid (6.3,1);
	\end{tikzpicture}
	\caption{Spectral representation of the considered scenario. The incumbent and secondary systems coexist in the same spectral band, and each one is assigned a different subset of subcarriers. 
		}
	\label{fig:spec_view}
	\vspace{-10pt}
\end{figure}
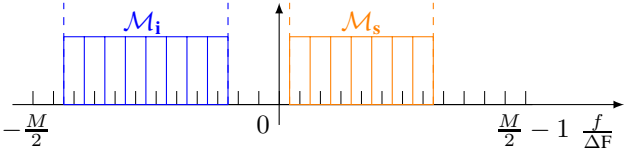

We consider a coexistence scenario where an incumbent CP-OFDM based system and an OFDM/OQAM secondary system share the same spectral band. The incumbent is assigned a set of active subcarriers $\mathcal{M}_\textbf{i}$ and the secondary a set $\mathcal{M}_\textbf{s}$. We assume that both systems use the same subcarrier spacing $\Delta\text{F}$ and time-symbol $T=\frac{1}{\Delta \text{F}}$. To focus on cross-interference between users, we consider a transmission on an additive white gaussian noise (AWGN) channel and do not take into account any pathloss or shadowing effect. The corresponding system model is represented in Fig.~\ref{fig:scenario} and an example of subcarriers distribution is given in Fig.~\ref{fig:spec_view}. 

Note that, strictly speaking, each interfering signal arrives at the receiver it interferes on with a certain delay. For example, naming $\delta_{\text{t},\textbf{s}}$ the propagation delay between the secondary transmitter and the incumbent receiver, the interfering signal received at the incumbent is not ${s}_\textbf{s}(t)$ but ${s}_\textbf{s}(t - \delta_{\text{t},\textbf{s}})$. However, we showed through simulation in \cite{BodinierICC2016} that this propagation delay has only little impact on the interference between OFDM/OQAM and CP-OFDM users. Without loss of generality, we will therefore neglect the propagation delays in our analysis.

Now that the system model is laid out, we present in the following Section a short background on CP-OFDM and OFDM/OQAM signal models.

\subsection{CP-OFDM Incumbent System}
We consider a CP-OFDM incumbent system with $M$ subcarriers, time-symbol $T$ and a CP duration of $T_\text{CP}$. As previously mentioned, it has a set of active subcarriers $\mathcal{M}_\textbf{i}$. The CP-OFDM time-domain signal transmitted on each active subcarrier $m_\textbf{i} \in \mathcal{M}_\textbf{i}$ is expressed as 
\begin{align}
s_{m_\textbf{i}}(t) =&
\sum\limits_{n_\textbf{i} \in \mathbb{Z}}\mathbf{d}_{m_\textbf{i}}[n_\textbf{i}]f_{\text{T},\textbf{i}}\left(t-n_\textbf{i}(T+T_\text{CP})\right)e^{j2\pi m_\textbf{i}\frac{t-n_\textbf{i}T_\text{CP}}{T}},\label{eq:smi}
\end{align}
with $\mathbf{d}_{m_\textbf{i}}$ the data vector of quadrature amplitude modulated (QAM) symbols transmitted on subcarrier $m_\textbf{i}$ and $f_{\text{T},\textbf{i}}$ the CP-OFDM transmit filter defined as:
\begin{align}
f_{\text{T},\textbf{i}}(t) &= \begin{cases}
\frac{1}{\sqrt{T}},& t \in [-T_\text{CP},T]\\
0,& \text{elsewhere.}
\end{cases}\label{eq:fTi}
\end{align}
The total transmitted signal is expressed as 
\begin{equation}
s_\textbf{i}(t) = \sum\limits_{m_\textbf{i}\in\mathcal{M}_\textbf{i}}s_{m_\textbf{i}}(t), \forall t \in \mathbb{R}.
\end{equation}

In this study, we consider an AWGN interference channel. Therefore, the signal at the input antenna of the CP-OFDM incumbent receiver is expressed as
\begin{equation}
y_\textbf{i}(t) = s_{\textbf{i}}(t)+s_{\textbf{s}}(t)
+w_\textbf{i}(t), \forall t \in \mathbb{R}\label{eq:yi}
\end{equation}
where $s_\textbf{s}(t)$ is the interfering OFDM/OQAM secondary signal whose expression will be detailed in the next section and $w_\textbf{i}(t)$ is the AWGN seen at the incumbent receiver.

Assuming perfect synchronization between the CP-OFDM transmitter and receiver, the $n_{\textbf{i}}$-th demodulated symbol on the $m_{\textbf{i}}$-th subcarrier of the CP-OFDM receiver is expressed, $\forall{n_\textbf{i}} \in \mathbb{Z}$, as 
\begin{align}
\mathbf{\hat{d}}_{m_\textbf{i}}[n_{\textbf{i}}] = \mathbf{{d}}_{m_\textbf{i}}[n_{\textbf{i}}] + \sum\limits_{m_\textbf{s}\in \mathcal{M}_\textbf{s}}\mathbf{\eta}_{m_\textbf{s}\rightarrow m_\textbf{i}}[n_\textbf{i}] + \mathbf{w}_{\textbf{i}}[n_\textbf{i}],
\end{align}
where $\mathbf{w}_\textbf{i}$ is the filtered white gaussian noise component expressed as 
\begin{equation}
\mathbf{w}_\textbf{i}[n_\textbf{i}] = \int\limits_{-\infty}^{\infty} f_{\text{R},\textbf{i}}(t-n_\textbf{i}(T+T_\text{CP}))e^{-j2\pi m_\textbf{i}\frac{t-n_\textbf{i}T_\text{CP}}{T}}{w_\textbf{i}(t)}dt, \forall{n_\textbf{i}} \in \mathbb{Z}\label{eq:wi}
\end{equation}
and $\mathbf{\eta}_{m_\textbf{s}\rightarrow m_\textbf{i}}$ is the interference injected by the $m_\textbf{s}$-th subcarrier of the secondary onto the $m_\textbf{i}$-th subcarrier of the incumbent, which is expressed $\forall{n_\textbf{i}} \in \mathbb{Z}$ as
\begin{equation}
\mathbf{\eta}_{m_\textbf{s}\rightarrow m_\textbf{i}}[n_\textbf{i}] = \int\limits_{-\infty}^{\infty}f_{\text{R},\textbf{i}}(t-n_\textbf{i}(T+T_\text{CP})) e^{-j2\pi m_\textbf{i}\frac{t-n_\textbf{i}T_\text{CP}}{T}} s_{m_\textbf{s}}(t)dt.\label{eq:etastoi}
\end{equation}
In \eqref{eq:wi} and \eqref{eq:etastoi}, $f_{\text{R},\textbf{i}}$ is the receive filter of the incumbent and is expressed as
\begin{align}
f_{\text{R},\textbf{i}}(t) &= \begin{cases}
\frac{1}{\sqrt{T}},& t \in [0,T]\\
0,& \text{elsewhere.}
\end{cases}\label{eq:fRi}
\end{align}

\subsection{Secondary OFDM/OQAM System}

We consider an OFDM/OQAM secondary system with $M$ subcarriers and time-symbol $T$, and name $\mathcal{M}_\textbf{s}$ its set of active subcarriers. The time-domain signal transmitted on each active subcarrier $ m_\textbf{s} \in \mathcal{M}_\textbf{s}$ of the secondary OFDM/OQAM system is expressed as \cite{CausICASSP2016}
\begin{equation}
s_{m_\textbf{s}}(t) =
\sum\limits_{n_\textbf{s} \in \mathbb{Z}}(-1)^{m_\textbf{s}n_\textbf{s}}\mathbf{d}_{m_\textbf{s}}[n_\textbf{s}]\mathbf{\theta}_{m_\textbf{s}}[n_\textbf{s}]f_{\text{T},\textbf{s}}\left(t-n_\textbf{s}\frac{T}{2}\right)e^{j2\pi m_\textbf{s}\frac{t}{T}}
\label{eq:sms}
\end{equation}
where $\mathbf{d}_{m_\textbf{s}}$ is the data vector of pulse amplitude modulated (PAM) real symbols and $\mathbf{\theta}_{m_\textbf{s}}[n_\textbf{s}]$ is a phase factor added to the symbols and defined as \cite{CausICASSP2016}
\begin{equation}
\mathbf{\theta}_{m_\textbf{s}}[n_\textbf{s}] = e^{j\frac{\pi}{2}\lfloor \frac{n_\textbf{s}+m_\textbf{s}}{2}\rfloor}.
\end{equation}
Besides, $f_{\text{T},\textbf{s}}$ is the transmit filter of the secondary system expressed as
\begin{align}
f_{\text{T},\textbf{s}}(t) &= \begin{cases}
g(t) & t \in [-\frac{KT}{2},\frac{KT}{2}]\\
0,& \text{elsewhere}
\end{cases}\label{eq:fTs}
\end{align}
where $K$ is called the overlapping factor, and $g$ is the used prototype filter. In the remainder of this study, we take $g$ as the PHYDYAS prototype filter with overlapping factor $K=4$, \cite{Bellanger}, defined as
\begin{equation}
g(t) = \sum\limits_{k=-K+1}^{K-1}\frac{G_{|k|}}{K} e^{j2\pi \frac{kt}{KT}}, t \in [-\frac{KT}{2},\frac{KT}{2}],\label{eq:g}
\end{equation}
with  $G_0 = 1$, $G_1 = 0.971960$, $G_2 = \frac{1}{\sqrt{2}}$, $G_3 = 0.235147$. Note that $g$ is a real and symmetric filter, such that $g^*(-t) = g(t), \forall t \in \mathbb{R}$.

The signal received at the input antenna of the OFDM/OQAM receiver is expressed in a similar manner as in \eqref{eq:yi} and is given by
\begin{equation}
y_\textbf{s}(t) = s_{\textbf{s}}(t)+s_{\textbf{i}}(t)
+w_\textbf{s}(t)
, \forall t \in \mathbb{R}\label{eq:ys}
\end{equation}

At the OFDM/OQAM receiver, the received signal is passed through the receive filter $f_{\text{R},\textbf{i}}$ expressed as
\begin{align}
f_{\text{R},\textbf{s}}(t) &= \begin{cases}
g^*(-t)=g(t) & t \in [-\frac{KT}{2},\frac{KT}{2}]\\
0,& \text{elsewhere.}
\end{cases}\label{eq:fRs}
\end{align}
Then, the real part of the filtered signal is taken to remove purely imaginary intrinsic interference \cite{CausICASSP2016}. Therefore, the $n_{\textbf{s}}$-th demodulated symbol on the $m_{\textbf{s}}$-th subcarrier of the OFDM/OQAM secondary receiver is expressed as 
\begin{align}
\mathbf{\hat{d}}_{m_\textbf{s}}[n_{\textbf{s}}] = \mathbf{{d}}_{m_\textbf{s}}[n_{\textbf{s}}] + \sum\limits_{m_\textbf{i}\in \mathcal{M}_\textbf{i}}\mathbf{\eta}_{m_\textbf{i}\rightarrow m_\textbf{s}}[n_\textbf{s}] + \mathbf{w}_{\textbf{s}}[n_\textbf{s}],
\end{align}
with 
\begin{flalign}
\mathbf{w}_\textbf{s}[n_\textbf{s}] &=\int\limits_{-\infty}^{\infty}\mathcal{R}\{ f_{\text{R},\textbf{s}}(t-n_\textbf{s}\frac{T}{2}) e^{-j2\pi m_\textbf{s}\frac{t}{T}}(-1)^{m_\textbf{s}n_\textbf{s}}{w_\textbf{s}(t)}\}dt\label{eq:ws}\\
\mathbf{\eta}_{m_\textbf{i}\rightarrow m_\textbf{s}}[n_\textbf{s}] &=\int\limits_{-\infty}^{\infty}\mathcal{R}\{ f_{\text{R},\textbf{s}}(t-n_\textbf{s}\frac{T}{2})& \nonumber\\&\times e^{-j2\pi m_\textbf{s}\frac{t}{T}}(-1)^{m_\textbf{s}n_\textbf{s}} s_{m_\textbf{i}}(t)\}dt\label{eq:etaitos}.
\end{flalign}

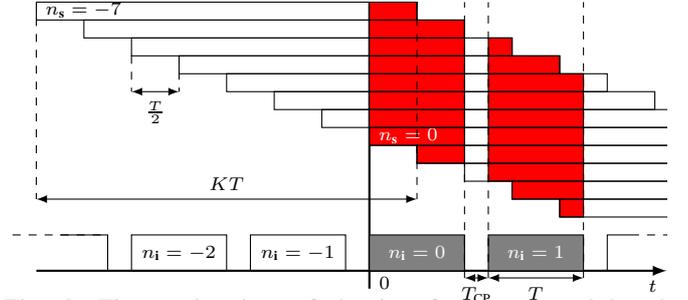
\begin{figure}[!t]
	\vspace{-5pt}
	\centering
	\begin{tikzpicture}[xscale=1.4,yscale=0.95]
	\clip (-15*\linewidth/40,4.25) rectangle (\linewidth*25/80,0);
	\draw[thick, ->,>=latex] (-7*\linewidth/20,0.5) -- node[at end, below left]{\scriptsize$t$} (\linewidth*25/80,0.5);
	\draw[thick] (0,0.25) -- node[pos=0.02,right]{\scriptsize$0$} (0,4.25);
	
	\draw[fill=red] (0,4.25) -- (\linewidth/20,4.25) -- (\linewidth/20, 4) -- (\linewidth/10,4) -- (\linewidth/10,2) -- (\linewidth/20,2) -- (\linewidth/20,2.25) -- (0,2.25) -- (0,4.25);
	\draw[fill=red] (\linewidth/8,3.75) -- (6*\linewidth/40,3.75) -- (6*\linewidth/40, 3.5) -- (8*\linewidth/40,3.5) -- (8*\linewidth/40,3.25) -- (9*\linewidth/40,3.25) -- (9*\linewidth/40,1.25) -- (8*\linewidth/40,1.25) -- (8*\linewidth/40,1.5) -- (6*\linewidth/40,1.5) -- (6*\linewidth/40,1.75)-- (5*\linewidth/40,1.75) -- (\linewidth/8,3.75);
	\draw [black] (-\linewidth/8,0.5) rectangle  node[text centered]{\scriptsize$n_\textbf{i}=-1$} (-\linewidth/40,1);
	\draw [black] (-\linewidth/4,0.5) rectangle  node[text centered]{\scriptsize$n_\textbf{i}=-2$} (-6*\linewidth/40,1);
	\draw [dashed] (-3\linewidth/8,1) --  (-11*\linewidth/40,1);
	\draw [black] (-13\linewidth/40,1) --  (-11*\linewidth/40,1) -- (-11*\linewidth/40,0.5);
	
	\draw [black, fill=gray] (0,0.5) rectangle  node[text centered]{\textcolor{white}{\scriptsize$n_\textbf{i}=0$}} (\linewidth*4/40,1);
	\draw [black, fill=gray] (\linewidth*1/8,0.5) rectangle  node[text centered,]{\textcolor{white}{\scriptsize$n_\textbf{i}=1$}} (\linewidth*9/40,1);
	\draw[dashed] (\linewidth*4/40,0.4) -- (\linewidth*4/40,4.25);
	\draw[dashed] (\linewidth*1/8,0.4) -- (\linewidth*1/8,4.25);
	\draw[dashed] (\linewidth*9/40,0.4) -- (\linewidth*9/40,4.25);
	
	\draw[<->,>=latex]  (\linewidth*4/40, 0.4) -- node[below]{\scriptsize$T_\text{CP} $} (\linewidth*1/8, 0.4);
	\draw[<->,>=latex]  (\linewidth*1/8, 0.4) -- node[below]{\scriptsize$T $} (\linewidth*9/40, 0.4);
	\draw (\linewidth*1/4,0.5) -- (\linewidth*1/4,1) -- (\linewidth*11/40,1);
	\draw[dashed] (\linewidth*11/40,1) -- (\linewidth*13/40,1);
	\draw [] (4*\linewidth/20,1.25) rectangle  (\linewidth*24/40,1.5) ;
	\draw [] (3*\linewidth/20,1.5) rectangle  (\linewidth*22/40,1.75) ;
	\draw [] (2*\linewidth/20,1.75) rectangle (\linewidth*20/40,2) ;
	\draw [] (\linewidth/20,2) rectangle  (\linewidth*18/40,2.25) ;
	\draw [] (0,2.25) rectangle (\linewidth*16/40,2.5) ;
	\node [above right] at (0,2.15) {\textcolor{white}{\scriptsize$n_\textbf{s}=0$}};
	\draw [] (-\linewidth/20,2.5) rectangle (\linewidth*14/40,2.75) ;
	\draw [] (-2*\linewidth/20,2.75) rectangle (\linewidth*12/40,3) ;
	\draw [] (-3*\linewidth/20,3) rectangle  (\linewidth*10/40,3.25) ;
	\draw [] (-4*\linewidth/20,3.25) rectangle  (\linewidth*8/40,3.5) ;
	\draw [] (-5*\linewidth/20,3.5) rectangle  (\linewidth*6/40,3.75) ;
	\draw [] (-6*\linewidth/20,3.75) rectangle (\linewidth*4/40,4) ;
	\draw [] (-7*\linewidth/20,4) rectangle  (\linewidth*2/40,4.25) ;
	\node [above right] at (-7*\linewidth/20,3.9) {\scriptsize$n_\textbf{s}=-7$};
	\draw[<->, >=latex] (-7*\linewidth/20,1.5) --  node[above]{\scriptsize$KT$} (\linewidth/20,1.5);
	\draw[<->, >=latex] (-5*\linewidth/20,3) --  node[below]{\scriptsize$\frac{T}{2}$} (-4\linewidth/20,3);
	\draw[dashed] (-5*\linewidth/20,3.5) -- (-5*\linewidth/20,3);
	\draw[dashed] (-4*\linewidth/20,3.5) -- (-4*\linewidth/20,3);
	\draw[dashed] (-7*\linewidth/20,4.25) -- (-7*\linewidth/20,1.5);
	\draw[dashed] (\linewidth/20,4.25) -- (\linewidth/20,1.5);
	\end{tikzpicture}
	\vspace{-10pt}
	\caption{Time axis view of the interference caused by the OFDM/OQAM transmission on CP-OFDM receiving windows $n_\textbf{i} = 0$ and $n_\textbf{i} = 1$. Both CP-OFDM symbols suffer the same amount of cross-interference.}
	\label{fig:interfering_symbols}
	\vspace{-15pt}
\end{figure}

	\section{Cross-Interference Analysis}
	\label{sec:analysis}
	
\begin{figure*}[!b]
	\normalsize
		\vspace*{-15pt}
	\hrulefill
	\vspace*{-5pt}
	\setcounter{tmpeqcnt}{\value{equation}}
	\setcounter{equation}{24}
	\begin{equation}
	\mathbf{\eta}_{m_\textbf{i}\rightarrow m_\textbf{s}}[n_\textbf{s}] = \frac{1}{\sqrt{T}}\sum_{n_\textbf{i}\in \mathbb{Z}}\mathcal{R}\bigg\{\mathbf{d}_{m_\textbf{i}}[n_\textbf{i}]e^{-j2\pi m_\textbf{i}\frac{n_\textbf{i}T_\text{CP}}{T}}(-1)^{m_\textbf{s}n_\textbf{s}}\int\limits_{n_\textbf{i}(T+T_\text{CP})-T_\text{CP}}^{n_\textbf{i}(T+T_\text{CP})+T_\text{CP}}g\left(t-n_\textbf{s}\frac{T}{2}\right)e^{j2\pi(m_\textbf{i}
		-m_\textbf{s})\frac{t}{T}}dt\bigg\}, \forall n_\textbf{s} \in \mathbb{Z}
	\label{eq:etaitosdev}
	\end{equation}
	\vspace*{-10pt}
	\setcounter{equation}{27}
	\begin{equation}
	I_{\textbf{i}\rightarrow\textbf{s}}(l) =  \frac{\sigma_{\mathbf{d}_\mathbf{i}}^2}{2K^2}(1+\frac{T_\text{CP}}{T})\sum\limits_{\tau=-KT+\frac{T}{2}}^{\frac{T}{2}}\left|\sum\limits_{k=-K+1}^{K-1}G_{|k|}e^{j\pi\frac{k}{K}\tau(1+\frac{T_\text{CP}}{T})}\text{sinc}\left(\pi\left(1+\frac{T_\text{CP}}{T}\right)\left(\frac{k}{K}+l\right)\right) \right|^2
	\label{eq:Iitos}
	\end{equation}
	\setcounter{equation}{\value{tmpeqcnt}}
	\vspace{-20pt}
\end{figure*}


\subsection{Cross-Interference at the Incumbent CP-OFDM Receiver}
Here, we derive the closed-form expression of interference seen at the incumbent CP-OFDM receiver as expressed in \eqref{eq:etastoi}.
Substituting the expression of $f_{\text{T},\textbf{s}}$ given by \eqref{eq:fTs} in \eqref{eq:sms}, and then putting both the resulting form of \eqref{eq:sms} and the expression of the incumbent receive filter $f_{\text{R},\textbf{i}}$ given by \eqref{eq:fRi} in \eqref{eq:etastoi}, we obtain the expression \eqref{eq:etastoidev} of $\mathbf{\eta}_{m_\textbf{s}\rightarrow{m_\textbf{i}}}[n_\textbf{i}]$. In the following, we focus on the mean interference power caused by subcarrier $m_\textbf{s}$ of the OFDM/OQAM secondary onto subcarrier $m_\textbf{i}$ of the CP-OFDM incumbent, which we define as
\addtocounter{equation}{1}
\begin{align}
\mathbf{I}_{m_\textbf{s}\rightarrow m_\textbf{i}}[n_\textbf{i}] &= E_{\mathbf{d_\mathbf{s}}}\{|\mathbf{\eta}_{m_\textbf{s}\rightarrow{m_\textbf{i}}}[n_\textbf{i}]|^2\}\label{eq:Imstomi}\\
&= \frac{\sigma_{\mathbf{d}_\textbf{s}}^2}{T}\sum\limits_{n_\textbf{s}\in \mathbb{Z}}\Bigg|\int\limits_{n_\textbf{i}(T+T_\text{CP})}^{n_\textbf{i}(T+T_\text{CP})+T}g\left(t-n_\textbf{s}\frac{T}{2}\right)\nonumber\\&\times e^{j2\pi (m_\textbf{s}
	-m_\textbf{i})\frac{t}{T}}dt\Bigg|^2,\label{eq:Imstomi2}
\end{align}
with $\sigma_{\mathbf{d}_\textbf{s}}^2$ the variance of the symbols modulated by the secondary OFDM/OQAM system. Note that this last expression is obtained by considering that the symbols $\mathbf{d}_\textbf{s}$ are independent and identically distributed (i.i.d.). Besides, note that the sum in $\eqref{eq:Imstomi2}$ can be reduced to the values of $n_\textbf{s}$ such that $g(t-n_\textbf{s}\frac{T}{2})$ is not null everywhere on the reception window corresponding to the $n_\textbf{i}$-th symbol of the receiver. This is shown in Fig.~\ref{fig:interfering_symbols},
where we also point out that the value of $n_\textbf{i}$ does not affect the mean interference power seen at the receiver. Note also that only the difference $m_\textbf{s}-m_\textbf{i}$ plays a role in \eqref{eq:Imstomi2}. Naming $l = m_\textbf{s}-m_\textbf{i}$ the spectral distance in terms of subcarriers, we therefore have
$\mathbf{I}_{m_\textbf{s}\rightarrow m_\textbf{i}}[n_\textbf{i}] = I_{\textbf{s}\rightarrow\textbf{i}}(l), \forall n_\textbf{i} \in \mathbb{Z}$.

After operating the change of variable $t\mapsto t-n_\textbf{i}(T+T_\text{CP})$ in \eqref{eq:Imstomi2},  we obtain, $\forall n_\textbf{i} \in \mathbb{Z}$,
\begin{align}
  \mathbf{I}_{m_\textbf{s}\rightarrow m_\textbf{i}}[n_\textbf{i}] &=\overset{I_{\textbf{s}\rightarrow\textbf{i}(l)}}{\overbrace{\frac{\sigma_{\mathbf{d}_\mathbf{s}}^2}{T}\sum\limits_{\tau=-KT+\frac{T}{2}}^{\frac{T}{2}}\underset{I_{\textbf{s}\rightarrow\textbf{i}}(l,\tau)}{\underbrace{\left|\int\limits_0^Tg(t-\tau)e^{j2\pi \frac{l t}{T}}dt\right|^2}}}}\label{eq:Iasasum}
\end{align} 
By including in \eqref{eq:Iasasum} the expression of $g$ given in \eqref{eq:g}, we obtain, $\forall l \in \mathbb{Z}$,
\begin{align}
I_{\textbf{s}\rightarrow\textbf{i}}(l,\tau) &= \frac{1}{K^2}\left|\sum\limits_{k=-K+1}^{K-1}G_{|k|} \int\limits_0^T e^{j2\pi(\frac{k(t+\tau)}{KT}+\frac{l t}{T})}dt\right|^2\\
 &= \frac{T}{K^2}\left|\sum\limits_{k=-K+1}^{K-1}G_{|k|}e^{j\pi\frac{k}{K}\tau}\text{sinc}(\pi(\frac{k}{K}+l)) \right|^2,
\end{align}
and the interference power injected by an OFDM/OQAM subcarrier to a CP-OFDM subcarrier at spectral distance $l$ is finally given by
\begin{align}
I_{\textbf{s}\rightarrow\textbf{i}}(l) = \frac{\sigma_{\mathbf{d}_\mathbf{s}}^2}{K^2}\sum\limits_{\tau=-KT+\frac{T}{2}}^{\frac{T}{2}}\Bigg|&\sum\limits_{k=-K+1}^{K-1}G_{|k|}e^{j\pi\frac{k}{K}\tau}\nonumber\\&\times \text{sinc}(\pi(\frac{k}{K}+l)) \Bigg|^2\label{eq:Istoi}
\end{align}

\subsection{Cross-Interference at the Secondary OFDM/OQAM Receiver}

We now focus on the cross-interference that is injected by the CP-OFDM incumbent transmitter onto the OFDM/OQAM receiver, as expressed in \eqref{eq:etaitos}. 
Substituting the expression of the CP-OFDM transmit filter \eqref{eq:fTi} into \eqref{eq:smi}, and then the resulting form of \eqref{eq:smi} and the expression of $f_{\text{R},\textbf{s}}$ given by \eqref{eq:fRs} into \eqref{eq:etaitos}, we obtain the expression of \eqref{eq:etaitosdev}, which is very close to \eqref{eq:etastoidev}, the only differences being the real part operator and some phase factors due to the OFDM/OQAM demodulation. In a similar way as \eqref{eq:Imstomi}, we define the mean interference power injected by subcarrier $m_\textbf{i}$ of the incumbent onto the $m_\textbf{s}$-th subcarrier of the secondary as
\addtocounter{equation}{1}
\begin{align}
\mathbf{I}_{m_\textbf{i}\rightarrow m_\textbf{s}}[n_\textbf{s}] &= E_{\mathbf{d}_\textbf{i}}\{|\mathbf{\eta}_{m_\textbf{i}\rightarrow m_\textbf{s}}[n_\textbf{s}]|^2\}\label{eq:Imitoms}\\
&=\frac{\sigma_{\mathbf{d}_\textbf{i}}^2}{2T}\sum_{n_\textbf{i}\in\mathbb{Z}}\Bigg|\int\limits_{n_\textbf{i}(T+T_\text{CP})-T_\text{CP}}^{n_\textbf{i}(T+T_\text{CP})+T}g\left(t-n_\textbf{s}\frac{T}{2}\right)\nonumber\\&\times e^{j2\pi(m_\textbf{i}
	-m_\textbf{s})\frac{t}{T}}dt\Bigg|^2,
\end{align}
with $\sigma_{\mathbf{d}_\textbf{i}}^2$ the variance of the symbols $\mathbf{d}_\textbf{i}$. Note that the factor $\frac{1}{2}$ comes from the fact that only the real part of the signal is taken at the OFDM/OQAM receiver.
The obtained expression is similar to \eqref{eq:Imstomi} and the different observations made in the previous section still hold true here. Therefore, following developments similar to \eqref{eq:Imstomi}-\eqref{eq:Istoi}, the power of interference injected by a CP-OFDM subcarrier to an OFDM/OQAM subcarrier at spectral distance $l$ is expressed as \eqref{eq:Iitos}.


Note that, if both systems transmit with the same energy per symbol $E_s$, $\frac{\sigma_{\mathbf{d}_\mathbf{i}}^2}{2} = \sigma_{\mathbf{d}_\mathbf{s}}^2 = \frac{E_s}{2}$ because each OQAM symbol transmits half the energy of a QAM symbol. Therefore, in the case where $T_\text{CP}=0$, which corresponds to a situation where the incumbent system does not use any CP, $I_{\textbf{i}\rightarrow\textbf{s}}(l) = I_{\textbf{s}\rightarrow\textbf{i}}(l),\forall l \in \mathbb{Z}$ and the incumbent and secondary systems interfere equally onto each other. 

From both \eqref{eq:Iitos} and \eqref{eq:Istoi}, it can be noticed that the rectangular time window of the CP-OFDM system incurs a sum of sine-cardinal in frequency, which slowly decrease and will therefore cause high interference to both systems, despite the well shaped prototype filter used by the OFDM/OQAM system. 

\subsection{Taking into Account Frequency Misalignments}
The closed forms we derived in the previous section were obtained in the case where the secondary and the incumbent systems agree on the exact same frequency basis. Therefore, the spectral distance $l$ between each subcarrier of the secondary and each subcarrier of the incumbent is an integer, i.e. $l \in \mathbb{Z}$. This is correct if the local oscillators (LO) of all users are perfectly synchronized. However, in a real setup, this is not the case. Indeed, LOs of mobile terminals have a typical accuracy of $\pm 1$ ppm with respect to their nominal frequency \cite{CompSyncConcepts}. At a carrier frequency of $2$ GHz, this can yield a misalignment between users of around $10^{4}$ Hz, which can become significant as it is close to the LTE subcarrier width of $15$ kHz.

\begin{figure}[!t]
	\centering
	\begin{tikzpicture}[scale=0.9]
	\draw[->,>=latex] (0,0) -- node[at end, below left]{$\frac{f}{\Delta \text{F}}$} (\linewidth,0);
	\draw[xstep=0.3,ystep=100,very thin] (0.3,0) grid (\linewidth*0.85,0.2);
	\node[below] at  (\linewidth*0.85, 0) {$\frac{M}{2}-1$};
	\node[below] at  (0.2, 0) {$-\frac{M}{2}$};
	\draw[->,>=latex] (\linewidth*0.437, -0.3) -- node[at start,left]{$0$} (\linewidth*0.437, 1.5);
	\draw[dashed,blue] (0.75, 0) -- (0.75, 1.5);
	\draw[dashed,blue] (3.15, 0) -- (3.15, 1.5);
	\draw[dashed,orange] (4.05, 0) -- (4.05, 1.5);
	\draw[dashed,orange] (6.15, 0) -- (6.15, 1.5);
	\node[blue] at (1.9,1.2) {$\mathcal{M}_\textbf{i}$}; 
	\node[orange] at (5.1,1.2) {$\mathcal{M}_\textbf{s}$};
	\draw[xstep=0.3, shift={(-0.15,0)},blue](0.9,0) grid (3.3,1);
	\draw[xstep=0.3, orange](4.2,0) grid (6.3,1);
	\draw[->,->=latex] (4.05,1) -- node[near end, above]{$\delta_\text{f}$} (4.2,1);
	\end{tikzpicture}
	\caption{Spectral representation of the considered scenario with frequency misalignment $\delta_{\text{f}}$ between coexisting systems. 
	}
	\label{fig:spec_view_df}
	\vspace{-15pt}
\end{figure}
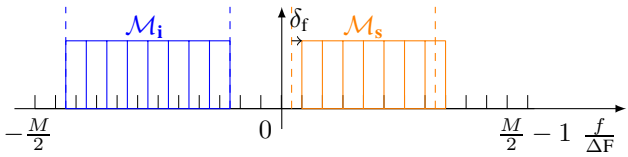

Here, we consider that the transmitter and receiver in each system achieve perfect frequency synchronization. However, the incumbent and secondary systems are not supposed to cooperate, and it is very likely that they will not be perfectly aligned in frequency.  This means that, taking the incumbent system as a reference, the $m_\textbf{s}$-th subcarrier of the secondary will not be modulated at base-band frequency $m_\textbf{s}\Delta\text{F}$ but at base-band frequency $(m_\textbf{s}+ \delta_\text{f})\Delta\text{F}$, where $\delta_\text{f} \in ]-0.5,\ 0.5]$ and is the frequency misalignment value between the secondary and incumbent systems. In other words, Fig.~\ref{fig:spec_view} is transformed into Fig.~\ref{fig:spec_view_df} where we see that the secondary transmission is misaligned with the band it should actually transmit in.

This simply leads to rewriting \eqref{eq:yi} and \eqref{eq:ys} as
\addtocounter{equation}{1}
\begin{align}
y_\textbf{i}(t) &= s_{\textbf{i}}(t)+s_{\textbf{s}}(t)e^{j2\pi\delta_\text{f}t}+w_\textbf{i}(t),\\
y_\textbf{s}(t) &= s_{\textbf{s}}(t)+s_{\textbf{i}}(t)e^{-j2\pi\delta_\text{f}t}+w_\textbf{s}(t).
\end{align}
Then, the analysis is led exactly as \eqref{eq:etastoi} - \eqref{eq:Iitos}. Redefining $l$ as $l= m_\textbf{s}+\delta_\text{f}-m_\textbf{i}$, mathematical derivations yield the exact same expression as \eqref{eq:Istoi} and \eqref{eq:Iitos}. Therefore, the closed forms we derived are also applicable in the presence of frequency misalignment between the incumbent and secondary systems i.e. $\forall l \in \mathbb{R}$.

	\section{Numerical Results}
	\label{sec:results}
\begin{figure}
	\vspace{-10pt}
	\centering
	\includegraphics[width=0.9\linewidth]{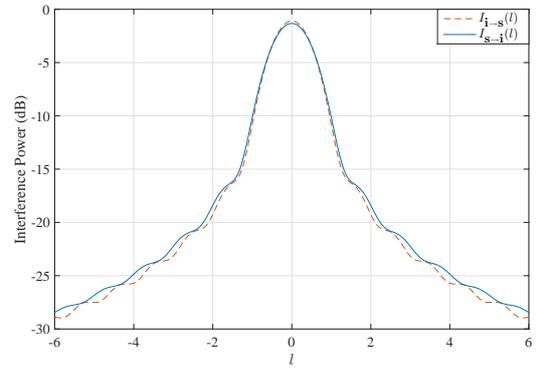}
	\caption{Values of cross interference injected at spectral distance $l$, according to  \eqref{eq:Iitos} and \eqref{eq:Istoi}.}
	\label{fig:Icontinuous}
	\vspace{-20pt}
\end{figure}

	In Fig.~\ref{fig:Icontinuous}, we represent the values of $I_{\textbf{s}\rightarrow\textbf{i}}(l)$ and $I_{\textbf{i}\rightarrow\textbf{s}}(l)$ according to the expressions of \eqref{eq:Iitos} and \eqref{eq:Istoi} for $T_\text{CP} = \frac{T}{8}$. We observe that both systems interfere almost equally onto each other. Note that, if we had set $T_\text{CP} = 0$, the two curves would perfectly overlap.

\begin{figure*}
	\vspace{-15pt}
	\captionsetup{justification=centering}
	\subfloat[Cross-interference seen at the Incumbent CP-OFDM receiver]{\includegraphics[width=0.45\linewidth]{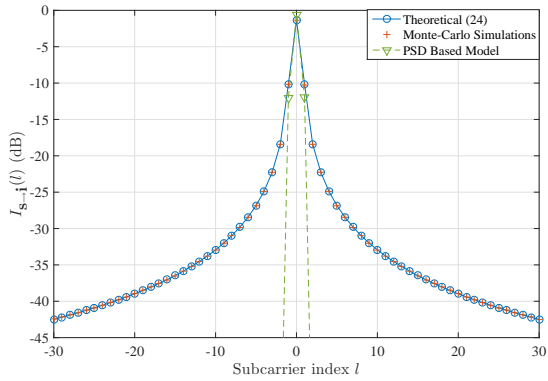}\label{fig:Istoi}}\hspace{0.1\linewidth}
	\subfloat[Cross-interference seen at the Secondary OFDM/OQAM receiver]{\includegraphics[width=0.45\linewidth]{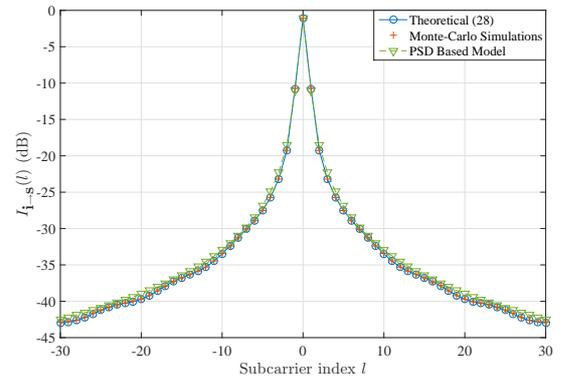}\label{fig:Iitos}}
	\caption{Cross-interference values for $\delta_\text{f} = 0$.\label{fig:I_deltaf0}}
	\vspace{-20pt}
\end{figure*}

In Fig.~\ref{fig:I_deltaf0}, the derived theoretical expressions are compared to results obtained through Monte-Carlo simulations. We represent the interference power seen at each subcarrier of index $l$ when the interferer transmits at subcarrier of index $0$. 
Parameters are set as follows: both the CP-OFDM incumbent and the OFDM/OQAM secondary systems have $M=512$ subcarriers. Both systems transmit symbols with unitary energy $\sigma_{\mathbf{d}_\mathbf{i}}^2 = 2\sigma_{\mathbf{d}_\mathbf{s}}^2=E_s=1$. Besides, we consider a CP of relative duration $T_\text{CP} = \frac{T}{8}$ and $\delta_\text{f} = 0$. Figures show that the expressions \eqref{eq:Istoi} and \eqref{eq:Iitos} perfectly match the simulation results.

Moreover, we show the results predicted by the PSD-based model used for example in \cite{skrzypczak2012ofdm,shaat2010computationally, Ihalainen2008}.
It is worth noticing that the PSD-based model gives a good approximation of the interference injected by the CP-OFDM incumbent onto the OFDM/OQAM secondary. However, it completely fails in estimating the interference injected by the OFDM/OQAM secondary onto the CP-OFDM incumbent. As pointed out in \cite{Bodinier2016ICT}, this is due to the fact that the PSD-based model does not take into account the rectangular receive window of the CP-OFDM receiver. Because the receive window of the OFDM/OQAM receiver is larger than the transmit filter of the CP-OFDM, the PSD appears to be a good measure of $I_{\textbf{i}\rightarrow\textbf{s}}$. On the contrary, as shown in Fig.~\ref{fig:interfering_symbols}, the CP-OFDM receive window truncates each OFDM/OQAM symbol and the good PSD properties of the latter are lost in the process.

\begin{figure}[!t]
	\centering
	\includegraphics[width=0.9\linewidth]{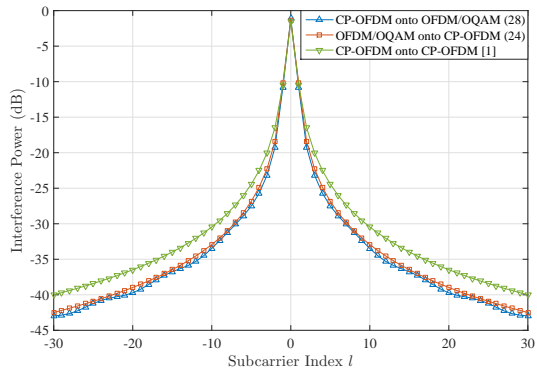}
	\caption{Comparison of cross-interference values in the case where the secondary uses CP-OFDM or OFDM/OQAM.}
	\label{fig:comp_OFDM}
	\vspace{-15pt}
\end{figure}

Finally, we compare the values of cross-interference in the considered scenario with those observed in a homogeneous scenario where both users would use CP-OFDM.
To do so, in Fig.~\ref{fig:comp_OFDM}, we compare the expressions of \eqref{eq:Istoi} and \eqref{eq:Iitos} with the interference tables derived by Medjahdi et. al. in \cite{Medjahdi2011} in the case where both the secondary and the incumbent users use CP-OFDM. Note that the interference observed by the incumbent and secondary users when they both use CP-OFDM is, at maximum, only $3$ dB higher to that observed when the secondary uses OFDM/OQAM. In real-world transmission, when the hardware impairments and the high power amplifier non-linearities come into play, this $3$ dB gap will fade out and the benefits of using OFDM/OQAM for coexistence with CP-OFDM systems are likely to disappear.

	\section{Conclusion}
	\label{sec:ccl}
	In this paper, we were able to develop exact closed forms of the cross-interference between coexisting OFDM/OQAM and CP-OFDM systems. As OFDM/OQAM is seen as a potential contender for certain 5G applications, the analysis led in this paper can be useful to dimension networks in scenarios where secondary 5G devices coexist with incumbent legacy users. 
Besides, the presented analysis can be generalized in the case where the secondary system would use other waveforms studied for 5G, such as filtered OFDM (f-OFDM) or Universal Filtered OFDM (UF-OFDM). Indeed, our analysis showed that the cross-interference is mainly caused by the rectangular window of the incumbent system. In other words, results presented in this paper show that coexistence with legacy CP-OFDM systems cannot be drastically improved by designing enhanced waveforms  only, but that it is necessary to modify the CP-OFDM receiver itself.

Hence, the derived closed forms show that the PSD-based model, widely used in the literature, is not fit to study coexistence scenarios, as it does not take into account the receive operations of the terminal that suffers from interference. This observation has a wide array of consequences: indeed, it shows that all the studies which rely on OOB emissions or PSD to rate cross-interference between users are inherently flawed, especially in the context of coexistence between two systems with different waveforms. However, a large number of studies recently released on that matter rely on these two measurements to rate the performance of various waveforms.

Moreover, standards like LTE-A usually define spectral masks that systems should respect to coexist smoothly. However, as these masks are based on PSD measurements as well, our results show that they are not adapted to the management and dimensioning of 5G systems.
Therefore, the results we derived in a particular scenario in this paper have broader applications and consequences. Especially, it is utterly important to rethink the models used to ensure coexistence between different physical layers. Future work will therefore focus on this particular aspect.

	
	\section*{Acknowledgement}
	\small This work was partially funded through French National Research Agency (ANR) project ACCENT5 with grant agreement code: ANR-14-CE28-0026-02.

	\normalsize
	
	
	
	%
	\normalsize
	\bibliographystyle{IEEEtran}
	\balance
	\bibliography{IEEEabrv,library}
\end{document}